\documentclass[aps,prb,preprint]{revtex4}
\usepackage{graphicx}
\usepackage{color}
\usepackage{amsmath,amssymb}
\usepackage{appendix}
\usepackage{bm}
\usepackage[pdffitwindow,colorlinks,citecolor={red},linkcolor={blue}]{hyperref}
\usepackage[capitalize]{cleveref}
\crefname{section}{Sec.}{Secs.}
\Crefname{section}{Section}{Sections}

\begin{document}

\title{Analytical Theory of Second Harmonic Generation from a
  Nanoparticle with a Non-Centrosymmetric Geometry}

\author{Raksha Singla}

\author{W. Luis Moch\'an}
\affiliation{Instituto de Ciencias F\'isicas, Universidad Nacional
  Aut\'onoma de M\'exico, Apartado Postal 48-3, 62251 Cuernavaca,
  Morelos, M\'exico}

\begin{abstract}

We analytically investigate the effect of a non-centrosymmetric
geometry in the optical second harmonic (SH) generation from a particle
made of a centrosymmetric material, in the interior of which quadratic
optical processes are suppressed. We consider a cylindrical particle
with a cross-section that is slightly deformed away from a circle and
with a radius much smaller than the wavelength. We
calculate the induced linear and nonlinear fields perturbatively in
terms of the deformation parameter and obtain the nonlinear dipolar and
quadrupolar hyperpolarizabilities, whose spectra we evaluate for
metallic and dielectric materials. We show that for very small
deformations the dipolar contribution to the response competes with
the quadrupolar term, and may even be dominant. We explore the
spectra of the hyperpolarizability and identify the contributions to
its structure for metallic and dielectric particles. We also discuss
the nature of SH radiation at various frequencies and find that it may
be dominated by the dipolar or the quadrupolar term, or that both may
compete yielding non-symmetric radiation patterns. Our calculation may
be employed to assess, calibrate and test numerical SH calculations.
\end{abstract}

\maketitle
\section{Introduction}

In recent years, the availability of novel techniques to produce
nanoparticles with different shapes has attracted much attention as
they can be engineered to exhibit unique nonlinear optical properties
which are sensitive to their environment as well as their
morphologies\cite{rev1nlo, rev2nlo}. Exploring second harmonic
generation (SHG) from nanoparticles and nanostructures has proven to
be an extraordinary tool to probe the properties of their
surfaces and interfaces. The surface sensitivity of the SH signal is
due to the fact that the bulk signal is strongly suppressed within
centrosymmetric systems; within the electric dipolar approximation,
SHG from a centrosymmetric system takes place only at the surface as
the bulk contribution is forbidden due to symmetry. The surface would
also be expected to dominate the SHG from nanoparticles made of
centrosymmetric materials, but if they possess a centrosymmetrical
shape there would be a cancellation of the
surface contributions to the nonlinear electrical dipole moment
arising from opposite sides, leaving only quadrupolar and higher
moments. However, when
subjected to a nonuniform polarizing field, a symmetrical particle can
generate SH due to the contributions arising from the excitation of a
nonlocal dipole moment\cite{brudny,huo,sun}. On the other hand, a
homogeneous external field would generate a nonvanishing dipolar SH
response even from nanoparticles made of centrosymmetric material if
their geometry is non-centrosymmetric. SHG from
nanoparticles, arrays of nanoparticles and nanostructured materials
with a variety of geometries have been demonstrated experimentally and
with supporting numerical
\cite{nanodimer,bowtie,butet,zhou,rectstrucmetal}
investigations. Several theoretical models have been reported in the
past to study SH scattering in the Rayleigh limit from small
symmetrical nanoparticles\cite{dadap,dadap2} or particles of arbitrary
size\cite{beer,fullwave} to include the effects of retardation and to
explore SHG within the framework of Mie
theory. An implementation of the discrete dipole approximation (DDA)
model to explore SHG from small nanoparticles of various kinds was
employed to study the influence of their shapes and sizes on their
nonlinear optical properties\cite{dda}. Different studies involving
various numerical computation techniques applied to diverse geometries
have since then been reported. Some of them used the finite-difference
time-domain (FDTD) method to
investigate the SHG from nanoholes in a metal
film\cite{rectstrucmetal}, and in particular, from an array of E
shaped nanoholes within a metal film\cite{zhou}. Others explored the
effect of deformations of metallic spheres\cite{bachelier} on SHG
using the finite element method (FEM). Investigation of SHG from gold split
ring resonators\cite{difftheomodel} used different theoretical models
and compared their applicability. A surface integral approach was used
to evaluate SH scattering from periodic metallic-dielectric
nanostructure\cite{sie3}, noble metal nanoparticles of arbitrary
shape\cite{sie,sie2}, and gold nanorod and
nanosphere\cite{mode}. Recently, a recursive method was employed to
study the SH susceptibility, the tuning
of its resonant structures and the corresponding nonlinear
polarization within metamaterials made of an array
of asymmetric cross shaped holes within metallic hosts \cite{ulises}.

As illustrated by the studies mentioned above, most of the efforts to
investigate SHG from nanoparticles have employed experimental or
{\em numerical} methods. To the best of our knowledge, there have been no
reports of an {\em analytical} calculation to study SHG from
nanoparticles with a non-centrosymmetric geometry. The purpose of this
paper is the calculation of the second order nonlinear response of an
isolated nanoparticle made up of a centrosymmetric material with small
deviations from a symmetrical geometry subjected to a homogeneous
external field,  obtaining and evaluating analytical expressions for
the nonlinear dipolar and quadrupolar hyperpolarizabilities. We
restrict our study to small deformations, which allows us to employ a
perturbative scheme in order to be able to solve the field equations
analytically for linear and SH induced fields within and beyond the
surface of the nanoparticle. We consider the simplest geometry for
which there is no centrosymmetry, namely, a cylinder having a
slightly deformed, near circular cross-section with a threefold
symmetry, and we use a nonretarded approximation to
obtain the near fields which we use afterwards to calculate the
electromagnetic fields in the radiation zone. We discuss the resonant
structure of the nonlinear response for a model metal and a dielectric
particle, the relation between the different
components of the response tensor owing to the symmetry in the system
and study the angular radiation patterns of the nanoparticle and their
evolution as the frequency sweeps across the various
resonances. As expected, we find the dipolar response to be highly
dependent on the deformation parameter as a deviation of only $1\%$
away from the symmetry of the shape of the particle results in a
strong competition with the quadrupolar response.

The structure of the paper is the following. In \cref{nldef}, we
describe our theory to investigate  analytically SHG from a
deformed cylindrical particle, obtaining expressions for the
nonlinear dipolar and quadrupolar hyperpolarizabilities (\cref{SHresp})
and the SH radiation patterns (\cref{SHrad}).
\Cref{results} illustrates our results for a deformed cylinder made
up of a Drude metal and a resonant dielectric. Finally, we present our
conclusions in \cref{conclusions}.


\section{Nonlinear response of a deformed cylinder}\label{nldef}

\subsection{Second order hyperpolarizabilities}\label{SHresp}

\begin{figure}
    \includegraphics[width=0.6\linewidth]{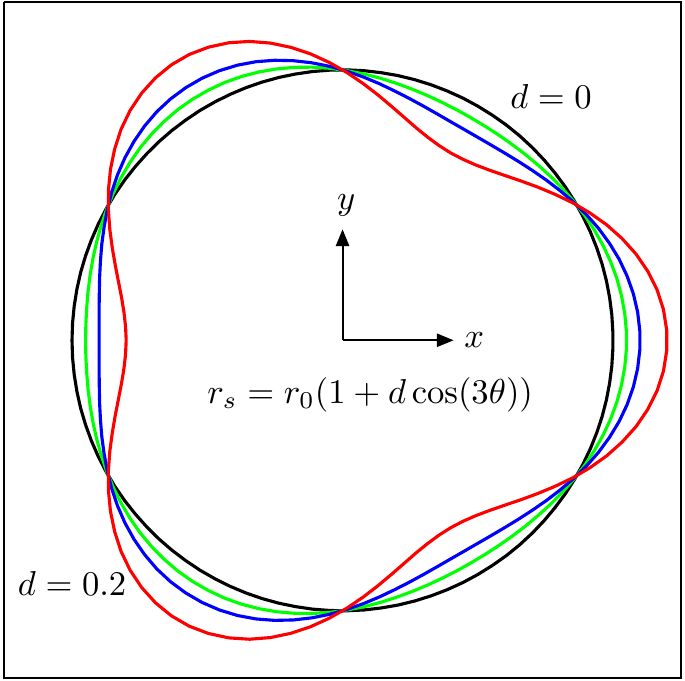}
    \caption{Cross-section of a deformed cylinder described by
      \cref{rtheta} for various values of the deformation parameter
      $d=0.0\ldots0.2$ (color online).}
    \label{fig1}
\end{figure}

We consider an isolated, infinitely long cylindrical particle placed
in vacuum with its axis along the ${\hat{\bm z}}$ direction and a slightly
deformed cross-section defined in polar $(r,\theta)$ coordinates as
\begin{equation}\label{rtheta}
  r_s(\theta)=r_{0}(1+d\cos3\theta)
\end{equation}
(See \cref{fig1}) where
$r_{0}$ is the radius of a symmetric nominal circular
cylinder and $d$ is a small deformation parameter. We remark that we
chose this geometry as it is the most simple one that lacks inversion
geometry. Also note that our system possesses a mirror $(y \to -y)$
and a $120^\circ$ rotational symmetry.
We will study the
nonlinear dipolar $\bm p$ and quadrupolar
$\bm Q$ moments induced per unit length along the axis of the cylinder. We
subject the particle to a homogeneous external electric
$\bm E^{\mathrm{ex}}$ field oscillating at frequency $\omega$.
Owing to the overall non-centrosymmetry of the
geometry of the particle, $\bm p$  has a
local contribution proportional to
$\bm E^{\mathrm{ex}}\bm E^{\mathrm{ex}}$ which we write as
\begin{equation}\label{alphad}
  p_i=\gamma^d_{ijk}E^{\mathrm{ex}}_jE^{\mathrm{ex}}_k
\end{equation}
where
$\gamma^d_{ijk}$ is the dipolar hyperpolarizbility and we use Einstein
summation convention. Similarly, the
induced quadratic 2D quadrupole moment, defined as
\begin{equation}\label{quad3}
  Q_{ij}=\int d^2r\,\rho(\bm r)(2r_ir_j-r^2\delta_{ij})
\end{equation}
(notice the difference with the usual 3D definition)
is given by
\begin{equation}\label{alphaq}
  Q_{ij}=\gamma^Q_{ijkl}E^{\mathrm{ex}}_kE^{\mathrm{ex}}_l
\end{equation}
where $\gamma^Q_{ijkl}$ is the quadrupolar hyperpolarizability. Here,
$\rho(\bm r)$ is the 2D charge density.

For
simplicity, we first assume that the external field is polarized along
the $\hat{\bm x}$ direction, $\bm E^{\mathrm{ex}}=E_0\hat{\bm
  x}$. Given the direction of
the external field and the symmetries in the system, the only nonzero
component of the nonlinear dipole moment induced in this case is
$p_{x}$, which we write as
\begin{equation}
  p_x=\gamma^d E_0^2.
  \label{px}
\end{equation}
in terms of a dipolar nonlinear response
$\gamma^d$ which is simply related all the components of the full dipolar
hyperpolarizability $\gamma^d_{ijk}$.
In this case the nonlinear quadrupole moment
has only two nonzero components
\begin{equation}
  Q_{xx}=-Q_{yy}=\gamma^Q E_0^2,
  \label{qxx}
\end{equation}
which we write in terms of a nonlinear
response $\gamma^Q$, related to the full quadrupolar
hyperpolarizability $\gamma^Q_{ijkl}$.
In this section we calculate analytically $\gamma^d$ and
$\gamma^Q$.

In the nonretarded regime, the linear self-consistent near field may be
obtained by solving Laplace's equation beyond and within the particle
and applying boundary conditions at its interface.
We  start with the general solution of Laplace's equation outside
\begin{equation}\label{phiout}
  \phi^{\mathrm{out}}_1 = \phi^{\mathrm{ex}}
  + \sum_{l=0}^{\infty}r^{-l}(s_l\cos l\theta+t_l\sin l\theta)
\end{equation}
and within
\begin{equation}\label{phiin}
   \phi^{\mathrm{in}}_1 = \sum_{l=0}^{\infty} r^{l}
     (u_l\cos l\theta+v_l \sin l\theta),
\end{equation}
the particle,
with multipolar coefficients $s_l$, $t_l$, $u_l$, and $v_l$ which we
expand as power series on the deformation parameter $d$
\begin{equation}\label{alphal}
  \beta_l=\sum_{n=0}^\infty \beta_l^{(n)} d^n,
\end{equation}
where $\phi^{\mathrm{ex}}=-E_0 r\cos\theta$  is the external scalar
potential and the generic coefficients $\beta_l$ stand for any of $s_l$, $t_l$,
$u_l$ or $v_l$. In order to perform analytical calculations, we will
restrict ourselves to small deformations and consider terms up to
linear order in $d$ only. Thus, solving Laplace's equation with
appropriate boundary conditions\cite{jackson} at the origin, the
surface of the particle, and infinity, we obtain the self-consistent
linear potential
\begin{equation}
  \frac{\phi^{\mathrm{out}}_1}{E_0}=r\cos\theta -
  \frac{1-\epsilon_1}{1+\epsilon_1}
  \frac{r_{0}^2}{r}\cos\theta  - d\left[\left(\frac{1-\epsilon_1}{1+\epsilon_1}
  \right)^2 \frac{r_{0}^3}{r^2}\cos2\theta
  +\frac{1-\epsilon_1}{1+\epsilon_1}
  \frac{r_{0}^5}{r^4}\cos 4\theta\right],
  \label{phi1out}
\end{equation}
\begin{equation}
  \frac{\phi^{\mathrm{in}}_1}{E_0}=\frac{2}{1+\epsilon_1}r\cos\theta +
  2d
  \frac{1-\epsilon_1}{(1+\epsilon_1)^2} \frac{r_{0}}{r^2} \cos 2\theta,
  \label{phi1in}
\end{equation}
where $\epsilon_g=\epsilon(g\omega)$ is the dielectric response of the
particle at $g$-th harmonic frequency, $g=1,2$. From these results we
may obtain the linear electric field $\bm E_1$.

The spatial variation of the self-consistent field within the particle
induces a nonlinear polarization\cite{jackson}
\begin{equation}\label{macp}
  \bm P^{nl}=n\bm p^{nl}
  -\frac{1}{2}n\nabla\cdot\bm q^{nl},
\end{equation}
which includes contributions from the nonlinear dipole $\bm p^{nl}$
and quadrupole $\bm q^{nl}$ moments of each
microscopic polarizable entity within the material, whose number density is
$n$. Note that $\bm q^{nl}$ may have a finite trace. Using the {\em dipolium}
model \cite{dipolium}, we write
\begin{equation}
  \bm p^{nl}=-\frac{1}{2e}\alpha_1\alpha_2\nabla E_1^2,
  \label{pnl}
\end{equation}
and
\begin{equation}
  \bm q^{nl}=-\frac{1}{e}\alpha_1^2\bm E_1\bm E_1.
  \label{qnl}
\end{equation}
in terms of the linear electric field and the linear polarizability
$\alpha_g\equiv\alpha(g\omega)$ evaluated at the fundamental ($g=1$) and
SH ($g=2$) frequencies, related to the dielectric function through
$\epsilon_g=1+4\pi n\alpha_g$.

The polarization, \cref{macp}, yields a nonlinear bulk charge
density given by
\begin{equation}\label{rho2}
  \rho^{nl}=-\nabla\cdot\bm P^{nl},
\end{equation}
which evaluates to zero up to linear order in $d$ (it would be nonzero
at order $d^2$). The termination of
the nonlinear polarization at the surface of the particle induces a
nonlinear surface charge with density
$\sigma^b=\bm P^{nl}(r_s^{-})\cdot\hat{\bm n}$ where
$\hat{\bm n}$
is a normalized outgoing vector perpendicular to the surface and
$r_s^-=r_s^-(\theta)$ denotes a position at the surface just inside the
particle. We employ the superindex $b$ to denote the bulk origin of
this surface charge. Using \cref{phi1out,phi1in,macp,pnl,qnl} we identify
\begin{equation}
  \sigma^b=4d\frac{n}{er_0}\frac{(1-\epsilon_1)}
        {(1+\epsilon_1)^3}
  \alpha_1(2\alpha_2-\alpha_1)\cos\theta E_0^2.
  \label{sigmanlb}
\end{equation}

As the inversion symmetry of the material is locally lost in a thin
selvedge region around the surface, there is a nonlinear polarization
induced at the surface of the particle which we write as
\begin{equation}
  P_i^s=\chi^{s}_{ijk} F_j F_k,
  \label{Ps}
\end{equation}
where ${\chi}^{s}_{ijk}$ are the components of the local nonlinear
surface susceptibility and the field $\bm F$ is defined in terms
of quantities that are continuous across the surface to avoid the
ambiguity about the position in the selvedge where the fields are to be
calculated; $\bm F$ is made up of the normal projection of the
displacement field and the parallel projection of the electric field
evaluated at the surface $r_s(\theta)$. Thus,
\begin{equation}
  \bm F(r_s)=\bm E_1(r_s^+)=\epsilon_1 \bm E_1^\perp(r_s^-)+
  \bm E_1^\parallel(r_s^-)),
  \label{fieldf}
\end{equation}
where $\bm E(r_s^-)=-\nabla \phi^{\mathrm{in}}(r_s)$ and
$\bm E(r_s^+)=-\nabla\phi^{\mathrm{out}}(r_s^+)$ and $\perp$ and
$\parallel$ denote the projections normal and parallel to the
surface.

We will assume that the thickness of the selvedge region is much
smaller than the radius of the cylinder and thus, that the surface can be
considered as locally flat. We will further assume {\em local} invariance
under rotations around the surface normal. Hence, the
surface susceptibility may be parametrized as
\begin{align}
  \chi^{s}_{ijk} = &\frac{(\epsilon_1-1)^2}{64\pi^2ne}\Big(\delta_{i\perp}
  \delta_{j\perp}\delta_{k\perp}\frac{a}{\epsilon_1^2}
  +[(1-\delta_{i\perp})
    (1-\delta_{j\perp})\delta_{k\perp}
    \nonumber \\
    &+(1-\delta_{i\perp})\delta_{j\perp}(1-\delta_{k\perp})]\frac{b}{\epsilon_1}
  +\delta_{i\perp}(1-\delta_{j\perp})(1-\delta_{k\perp})f\Big),
  \label{chidip}
\end{align}
in a local reference frame where one of the cartesian directions is
perpendicular and the others are parallel to the surface.
Here, $a$, $b$, and $f$ are dimensionless functions of $\omega$ used
to parametrize the response of the surface\cite{rudnick} given in the
{\em dipolium} model\cite{dipolium} by
\begin{equation}
  a(\omega)=2\frac{(\epsilon_2-\epsilon_1)
  (2\epsilon_1 - \epsilon_2 - \epsilon_1\epsilon_2)
    +\epsilon_1^2 (1-\epsilon_2) \log(\epsilon_1/\epsilon_2)}
  {(\epsilon_2-\epsilon_1)^2},
  \label{a}
\end{equation}
\begin{equation}
  b=-1,\label{b}
\end{equation}
\begin{equation}
  f= 0\label{f}.
\end{equation}

The normal component of the nonlinear polarization induced on the
surface of the cylinder is obtained by substituting
\cref{fieldf,chidip} in \cref{Ps},
\begin{align}
  P^s_\perp=&  \frac{1}{32\pi^2ne}
  \left(
     \frac{1-\epsilon_1}{1+\epsilon_1}
  \right)^2
  \bigg\{
    (a+f)+(a-f)\cos2\theta \nonumber\\
    &-d
    \left[
       \left(
          4\frac{1-\epsilon_1}{1+\epsilon_1}
          +3(a-f)
       \right)
       \cos\theta
          +4\frac{1-\epsilon_1}{1+\epsilon_1}\cos3\theta
          -3(a-f)\cos5\theta
          \right]
    \bigg\}
       E_0^2.
  \label{psperp}
\end{align}

The variation of the tangential component of the nonlinear surface
polarization along the surface yields another contribution to the
surface charge $\sigma^s$ beyond that due to the termination of the
bulk nonlinear polarization $\sigma^b$, where we use the superscript
$s$ to denote its surface origin. It is given by
\begin{equation}\label{sigmas}
  \sigma^s=-\nabla_{\parallel} \cdot \bm P^s_{\parallel},
\end{equation}
where
$\nabla_{\parallel}$ is the gradient operator projected along
the surface and
$\bm P^s_{\parallel}$ is the projection of $P^s$ along the
surface. Substituting \cref{Ps,chidip} in \cref{sigmas}
we obtain
\begin{align}
  \sigma^s=\frac{b}{8\pi^2ner_0}
  \left(\frac{1-\epsilon_1}{1+\epsilon_1}\right)^2
  \left[\cos2\theta
    +d\left(\cos\theta-6
    \frac{1-\epsilon_1}{1+\epsilon_1}\cos3\theta+7\cos5\theta\right)\right]
    E_0^2.
  \label{sigmanls}
\end{align}

The screened scalar potential $\phi_2$ induced at
the SH frequency has $\rho^{\mathrm{nl}}$ ($=0$) as an {\em external} bulk
source and the total nonlinear charges induced at the surface
$\sigma^b$ and $\sigma^s$ as external surface sources, together with the
normal polarization $P^s_\perp$, which are accounted through the
boundary conditions. The external sources have to be screened by the
linear response of the particle at SH frequency $\epsilon_{2}$. Thus
the equation to be solved for the quadratic scalar potential is
\begin{equation}\label{phi2}
\nabla^2\phi_2=
\begin{cases}
  0& \text{(outside)}\\
  -4\pi \rho^{\mathrm{nl}}/\epsilon_2=0& \text{(inside)}
\end{cases}
\end{equation}
subject to the boundary conditions
\begin{equation}
  \hat{\bm n}\cdot\nabla\phi_2(r_s^+)-\epsilon_2
    \hat{\bm n}\cdot\nabla\phi_2(r_s^-)=-4\pi
      (\sigma^b+\sigma^s)
      \label{bc22}
\end{equation}
\begin{equation}
  \phi_2(r_s^+)-\phi_2(r_s^-)=4\pi P^s_{\perp}
  \label{bc21}
\end{equation}
Eq.\,(\ref{bc22}) is the discontinuity of the normal
component of the displacement field due to the presence of the nonlinear
surface charge. Eq.\,(\ref{bc21}) is the discontinuity of the scalar
potential due to the presence of the normal nonlinear surface polarization
$P^{s}_{\perp}$
which is a dipole layer across the selvedge of the
particle. Solving Laplace's equation perturbatively to obtain
the self-consistent scalar potential at the SH frequency with terms
up to linear order in $d$ we obtain on the outside
\begin{align}
  \frac{\phi^{\mathrm{out}}_2}{E_0^2}=&
  \frac{d}{4 \pi ne}
  \frac{(1-\epsilon_1)^2}{(1+\epsilon_2)(1+\epsilon_1)^2}
  \left(4\frac{\epsilon_1-2\epsilon_2+1}{1+\epsilon_1}
  +
  2b\frac{1+3\epsilon_2}{1+\epsilon_2}
         +\frac{\epsilon_2}{2}\frac{(\epsilon_2-3)(a-f)}{1+\epsilon_2}
         \right.
         \nonumber\\
         +&\left.\frac{\epsilon_2}{2}
         \frac{(7\epsilon_1-1)f+(\epsilon_1-7)a}{1+\epsilon_1}\right)
         \frac{ r_0}{r}\cos\theta
         \nonumber\\
         +&\frac{1}{8\pi ne}\left(\frac{1-\epsilon_1}
       {1+\epsilon_1}\right)^2
       \frac{\epsilon_2(a-f)+2b}{1+\epsilon_2}\,
       \frac{r_0^2}{r^2}\cos2\theta
       +\ldots
         \label{phi2d}
\end{align}
where we only kept the dipolar and quadrupolar contributions, and
neglected higher multipoles, all of which are at least of order $d$,
as their contribution to the radiation fields would be insignificant
for small particles, at least by a factor of order $r_0/\lambda$, with
$\lambda$ the wavelength.

Finally, we compare Eq.\,(\ref{phi2d}) to the general expression of
the 2D scalar potential in polar coordinates, we identify the
corresponding components of the multipolar
moments, and  from \cref{px,qxx} we obtain the dipolar
\begin{align}
  \gamma^{d}=&\frac{d r_0}{4 \pi ne}\frac{(1-\epsilon_1)^2}
        {(1+\epsilon_2)(1+\epsilon_1)^2}
        \left(
          4\frac{\epsilon_1-2\epsilon_2+1}{1+\epsilon_1}
          \right.\nonumber\\
          &\left.+2b\frac{1+3\epsilon_2}{1+\epsilon_2}
          +\frac{\epsilon_2}{2}\frac{(\epsilon_2-3)(a-f)}{1+\epsilon_2}
          +\frac{\epsilon_2}{2}\frac{(7\epsilon_1-1)f+(\epsilon_1-7)a}
          {1+\epsilon_1}\right)
          \label{hyperdip}
\end{align}
and quadrupolar
\begin{equation}\label{hyperquad}
  \gamma^{Q}=\frac{r_0^2}{8\pi ne}\left(\frac{1-\epsilon_1}
        {1+\epsilon_1}\right)^2
        \frac{\epsilon_2(a-f)+2b} {1+\epsilon_2}
\end{equation}
nonlinear response functions.

To the lowest order in the deformation parameter, $\gamma^d$
is proportional to $d$, and thus, as expected, the dipolar response
would disappear for a centrosymmetric circular cross-section. On the
other hand, $\gamma^Q$ is a
constant, with no contribution proportional to $d$, and therefore it
coincides with the corresponding result for a cylinder of circular
cross-section. The contributions arising from the bulk
and the surface to the nonlinear hyperpolarizabilities can be easily
identified as the latter are proportional to the surface parameters
$a$, $b$, and $f$. Thus, $\gamma^d$ has both surface and bulk
contributions, while $\gamma^Q$ has only surface contributions. Both
$\gamma^d$ and $\gamma^Q$ inherit the spectral
structure
of the surface parameters, namely, of $a(\omega)$, and also exhibit
additional resonances corresponding to the excitation of surface
plasmons or surface plasmon-polaritons at the fundamental and second
harmonic frequencies, given by
$\epsilon_1=-1$ and $\epsilon_2=-1$ respectively.

\begin{figure}
  \includegraphics[width=0.7\linewidth]{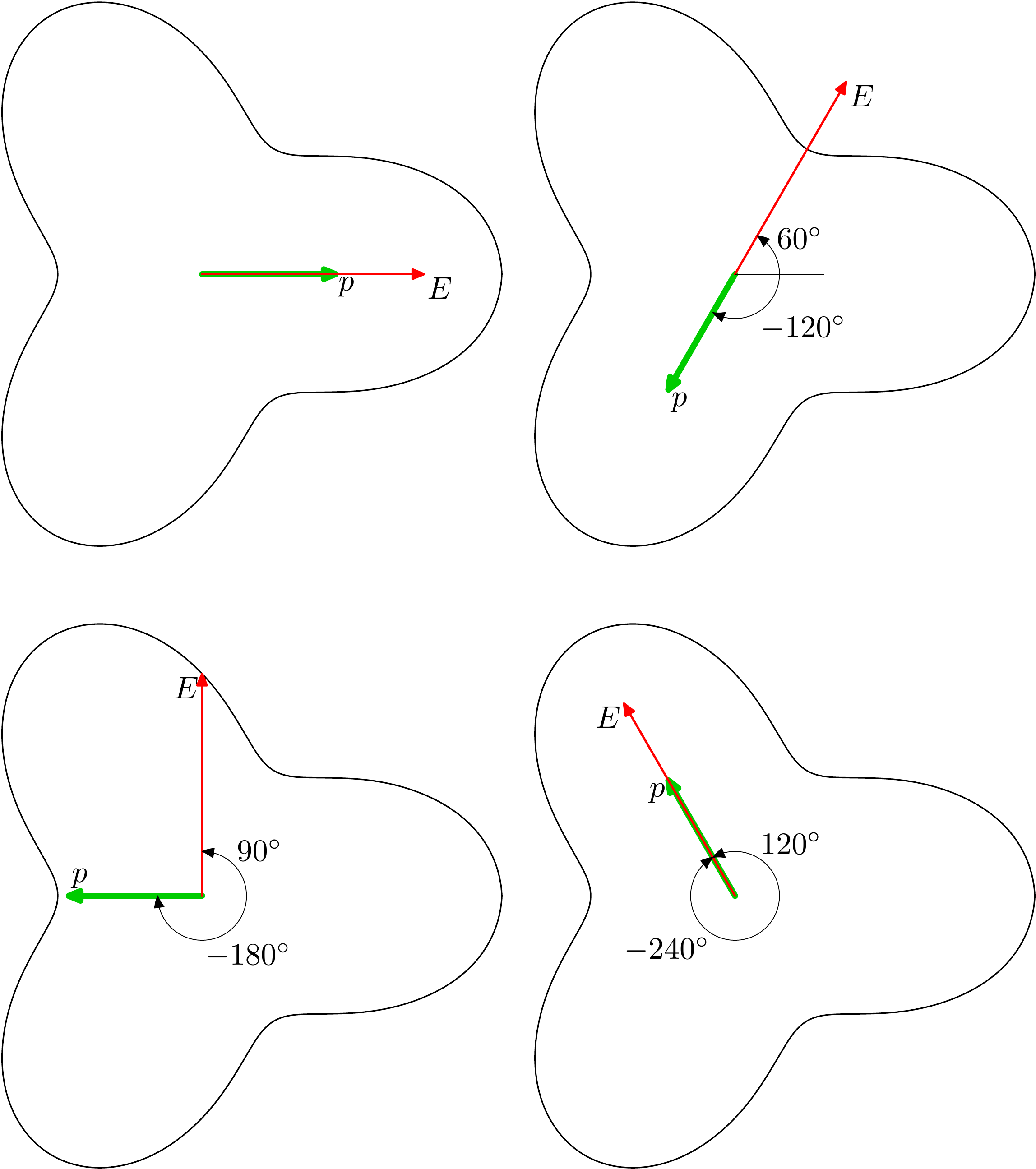}
  \caption{\label{sym}
    Direction of the quadratic dipole moment induced in a deformed
    cylindrical particle by an external field with different
    directions with respect to the horizontal. As the field rotates
    clockwise by an angle $\theta$ the dipole rotates counterclockwise
    by $2\theta$ (color online).}
\end{figure}

Above we have explicitly
shown the calculation of the nonlinear response of the particle
with the external field in the $\hat{\bm x}$ direction. One can
similarly evaluate the response of the particle to the external field
pointing in other directions.
However, due to the mirror and the $120^\circ$ rotation symmetry in our system,
all the in-plane components of the dipolar hyperpolarizability are
zero except for
$\gamma^d_{xyy}=\gamma^d_{yxy}=\gamma^d_{yyx}=-\gamma^d_{xxx}=-\gamma^d$. We
have verified these results by repeating the calculations above for
external fields pointing along different directions. It turns out that
given these symmetry related relations, the nonlinear dipole moment
induced in the deformed cylinder rotates anticlockwise by an angle
$2\theta$ when the external electric field is rotated clockwise by
$\theta$ (see \cref{sym}). The symmetry in our system
leads to an isotropic quadrupolar response given by
$Q_{ij}=\gamma^Q(2E_iE_j-E^2\delta_{ij})$, the quadratic
quadrupolar moment has a principal axis along the external field and
the only non-null components of the quadrupolar hyperpolarizability are
$\gamma^Q_{xxxx}=-\gamma^Q_{xxyy}=-\gamma^Q_{yyxx}=\gamma^Q_{yyyy}=\gamma_Q$ and
$\gamma^Q_{xyxy}=\gamma^Q_{yxxy}=\gamma^Q_{xyyx}=\gamma^Q_{yxyx}=2\gamma_Q$. Thus,
for this system our calculations of $\gamma^d$ and $\gamma^Q$ using
an external field along $\hat{\bm x}$ are sufficient to obtain the
full response response of the particle.

\subsection{SH radiation}\label{SHrad}

Now we turn our attention towards the calculation of the SH angular
radiation pattern. Following a procedure similar to the $3D$ case, one
can write down the expressions for the radiated electromagnetic fields
in $2D$ due to localized distributions of charges and
currents. Using the vector potential from \cref{dip,quad2} (see
Appendix), we calculate the radiated
electromagnetic fields.
\begin{equation}
  \bm B=(1+i)k^{3/2}\left((\hat{\bm r}\times \bm p)
    -\frac{i}{4}k^2(\hat{\bm r}\times(\bm Q\cdot\hat{\bm r}))\right)
    e^{ikr}\sqrt{\frac{\pi}{r}},
    \label{Brad}
\end{equation}
\begin{equation}
  \bm E=\bm B\times\hat{\bm r},
  \label{Erad}
\end{equation}
where $k$ is the wavenumber and $\hat{\bm r}$ is the unit vector in
the direction of 
observation. The time averaged power radiated per unit angle $\theta$
due to these radiated fields is
\begin{equation}
  \frac{dP}{d\theta}=\frac{rc}{8\pi}
  \operatorname{Re}[\bm E \times \bm B^{*}]\cdot \hat{\bm r}.
  \label{angrad}
\end{equation}
From \cref{hyperdip,hyperquad,Brad,Erad,angrad} we obtain
\begin{align}
  \frac{dP}{d\theta} =&\frac{cE_0^4k^3}{4}\Big(|\gamma^d|^2\sin^2\theta
  +4k\operatorname{Im}(\gamma^d\gamma^{Q*})\sin^2\theta\cos\theta \nonumber
  \\
  &+4k^2|\gamma^Q|^2\sin^2\theta\cos^2\theta\Big).
  \label{totrad}
\end{align}
The first and last terms correspond to dipolar and quadrupolar
radiation, while the middle term corresponds to their interference.


\section{Results}\label{results}

We consider a particle made up of a Drude metal characterized by its bulk
plasma frequency $\omega_p$ and electronic relaxation time $\tau$,
with dielectric function
$\epsilon(\omega)=1-\omega_p^2/(\omega^2+i\omega/\tau)$
\cite{ashcroft}, with a small deformation parameter $d=0.01$ and
with a small dissipation $1/\omega_p\tau=0.01$. We remark that while the
nonlinear dipolium model
\cite{dipolium} corresponds to a continuous distribution of small
polarizable entities, its results agree with those of a nonlinear
local {\em jellium} with a continuous electronic density profile at
the surface, and thus, it may be applied to metallic surfaces
\cite{hydro}.
\begin{figure}
  \includegraphics[width=.7\textwidth]{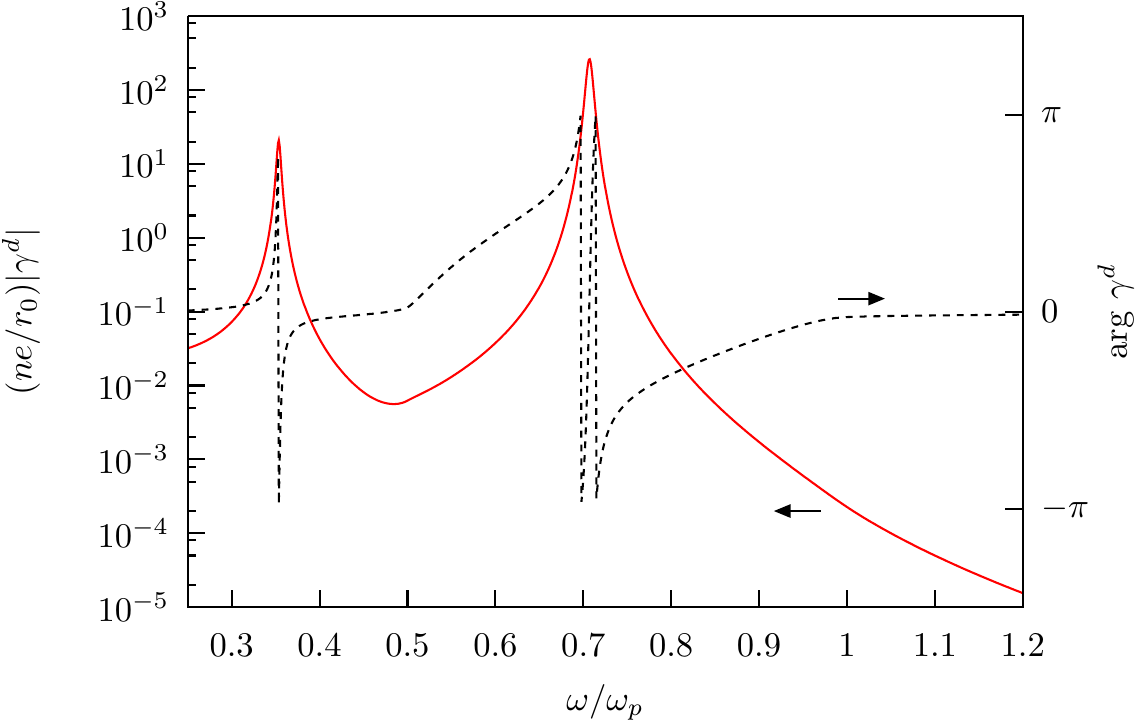} \\
  \includegraphics[width=.7\textwidth]{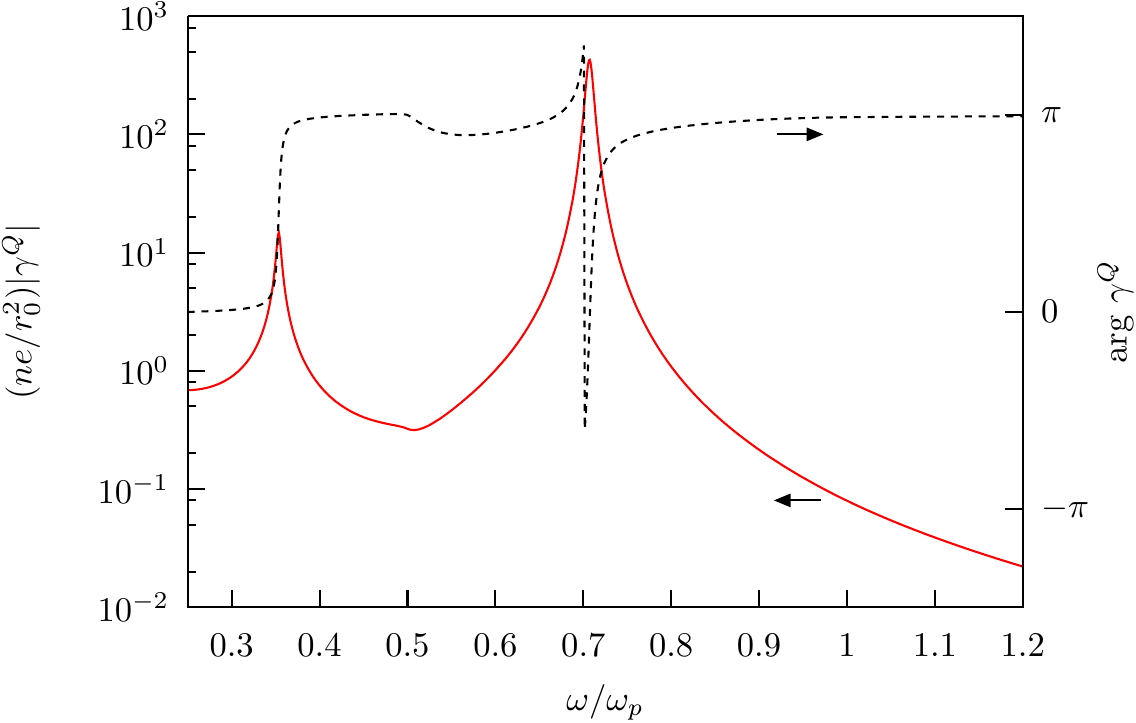}
  \caption{\label{drudegam} Normalized absolute value (solid) and
    phase (dashed) of the
    dipolar (upper
    panel) and quadrupolar (lower panel) nonlinear response functions
    $\gamma^d$ and $\gamma^Q$ for a cylinder with deformation
    parameter $d=0.01$  made of a Drude metal
    as a function of the normalized
    frequency $\omega/\omega_p$  (color online).}
\end{figure}
In \cref{drudegam} we show the absolute values and phases
of the nonlinear dipolar and quadrupolar response functions $\gamma^d$
and $\gamma^Q$. Notice that both display very large resonant peaks
corresponding to the surface plasmon resonance of the cylinder
$\omega_{sp}=\omega_p/\sqrt{2}$ and to its subharmonic.
Beyond abrupt changes at the resonances, the phase of $\gamma^d$
shifts away from 0 in a wide region that spans
from $\omega_p/2$ up to $\omega_p$. This is due to the logarithm term
in \cref{a}, whose argument changes sign as $\omega$ or $2\omega$
sweeps across the plasma frequency \cite{dipolium}. The phase of
$\gamma^Q$ also displays a smooth variation of around $2\pi$ in
the same region.

\begin{figure}
  \includegraphics[width=.7\textwidth]{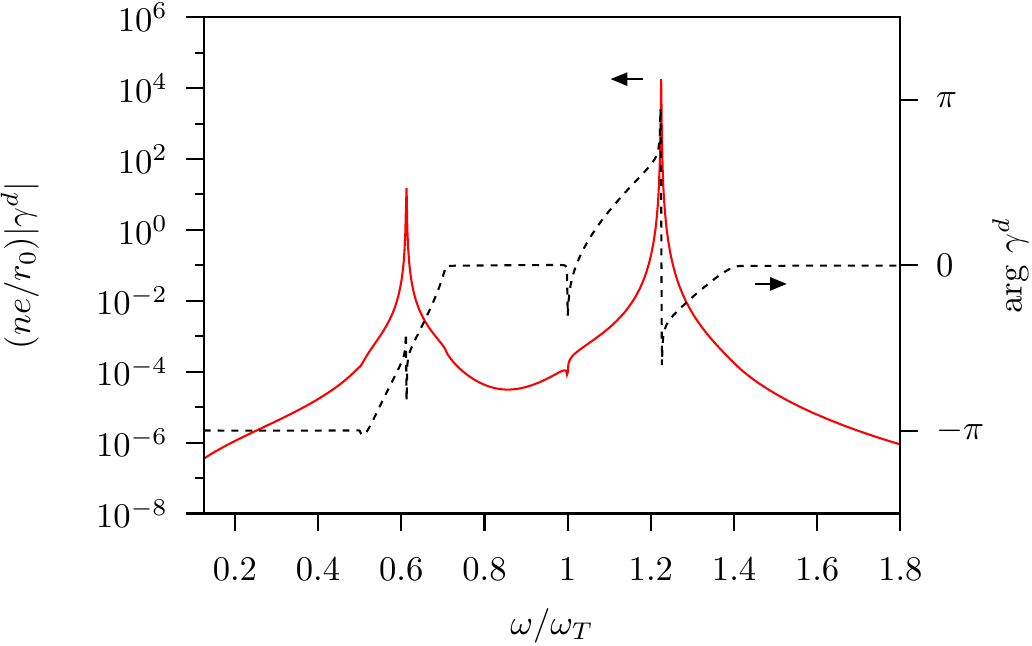} \\
  \includegraphics[width=.7\textwidth]{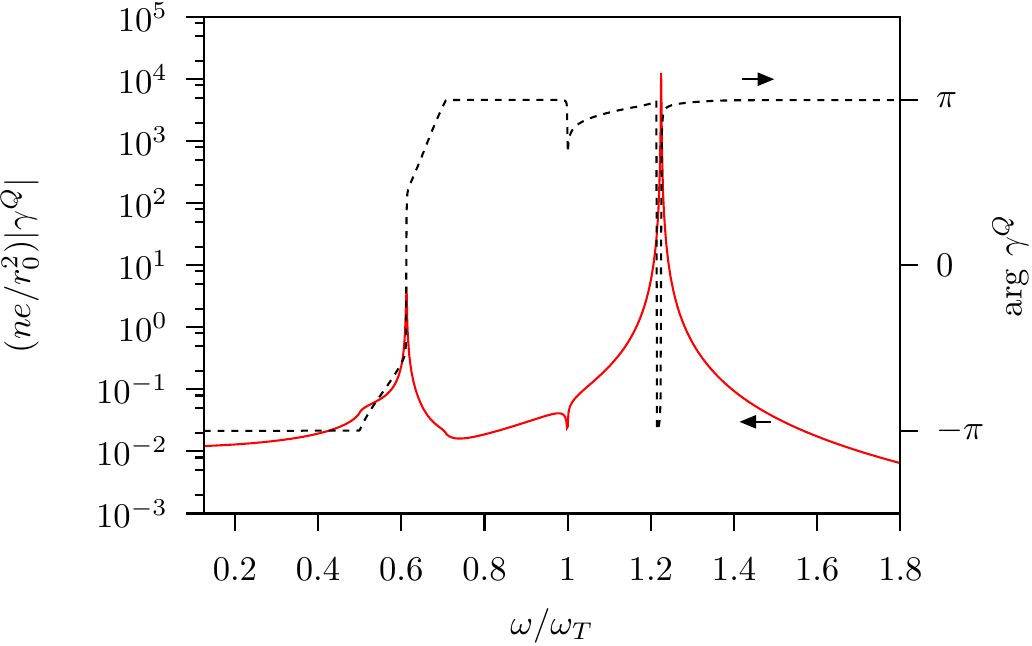}
  \caption{  \label{dielgam}
    Normalized absolute value (solid) and phase (dashed) of the dipolar (upper
    panel) and quadrupolar (lower panel) nonlinear response functions
    $\gamma^d$ and $\gamma^Q$ for a cylinder made of a dispersive
    dielectric with a Lorentzian resonance characterized by a
    longitudinal $\omega_L$ and a transverse $\omega_T$ frequency
    with $\omega_L=\sqrt{2}\omega_T$, and a
    deformation parameter $d=0.01$ as a function of the normalized
    frequency $\omega/\omega_T$ (color online).}
\end{figure}

In \cref{dielgam} we show the absolute values and phases of $\gamma^d$
and $\gamma^Q$ for a similar deformed cylinder but made up of an
insulator. We assume its dielectric function is dispersive and for
simplicity we assume it has a
single resonance described
by a simple Lorentzian\cite{ashcroft} form given by
$\epsilon(\omega) = (\omega_L^2-\omega^2-i\omega/\tau) /
(\omega_T^2-\omega^2-i\omega/\tau)$
 where
$\omega_L$ and $\omega_T$ are the frequencies of the longitudinal and
transverse optical modes respectively and included a very small
dissipation characterized by $\tau$. For definitiveness, we took
$\omega_L=\sqrt2\omega_T$.
Both $\gamma^d$ and $\gamma^Q$
show strong resonant peaks corresponding to the excitation of the
surface plasmon-polariton at $\omega=\omega_{spp}=\omega_T\sqrt{3/2}$
and its
subharmonic. The dielectric function crosses zero at
$\omega_L$ and has a pole at
$\omega_T=1$. The phase of both $\gamma^d$ and $\gamma^Q$ grows smoothly
between $\omega_T/2$ and $\omega_L/2$, and between $\omega_T$ and
$\omega_L$, save for abrupt jumps at $\omega_{spp}/2$, $\omega_{T}$ and
$\omega_{spp}$, and remain constant otherwise.   
Some
features in the phase of
$\gamma^d$ and $\gamma^Q$ are inherited
from those of the parameter $a$ \cite{dipolium}.

\begin{figure}
  \begin{center}
    \begin{tabular}{cc}
      \includegraphics[width=0.4\linewidth]{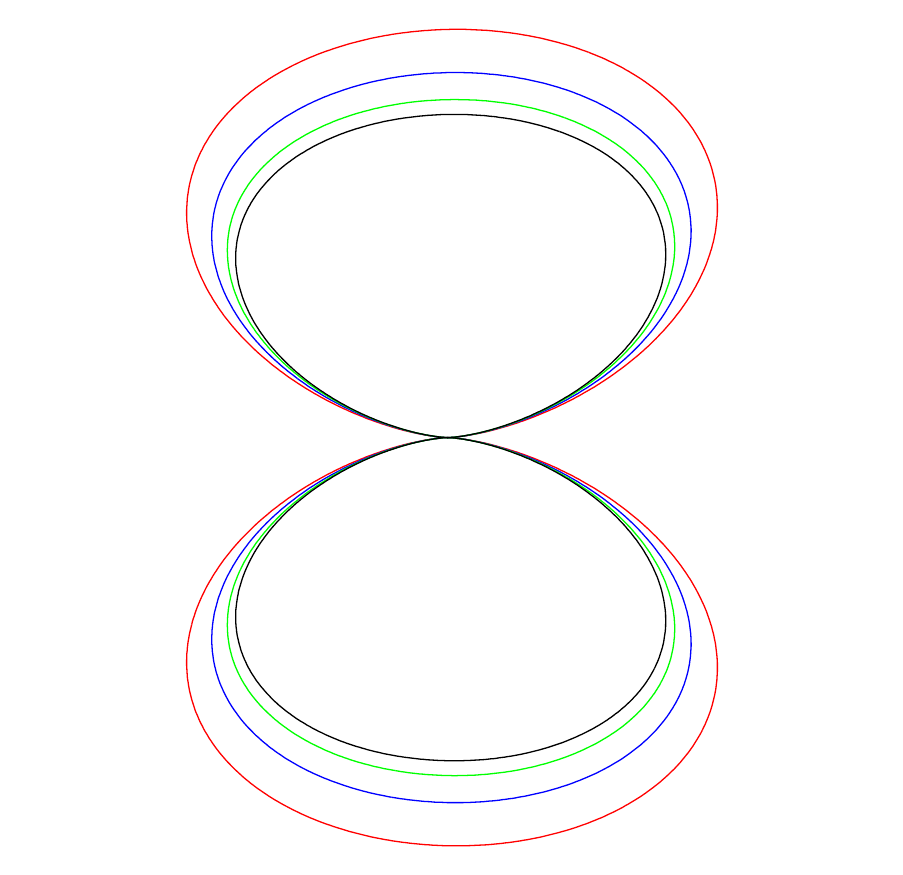} &
      \includegraphics[width=0.45\linewidth]{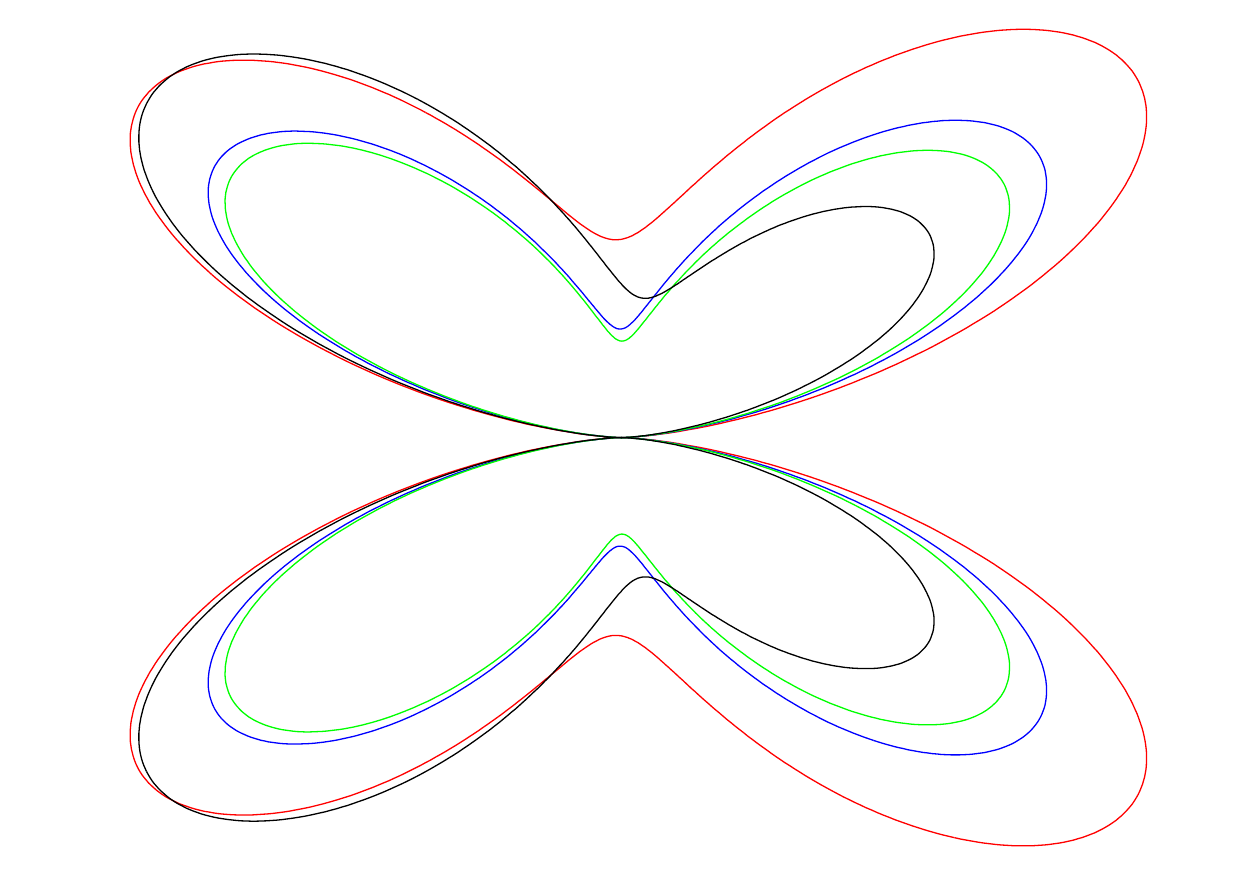} \\
      \includegraphics[width=0.4\linewidth]{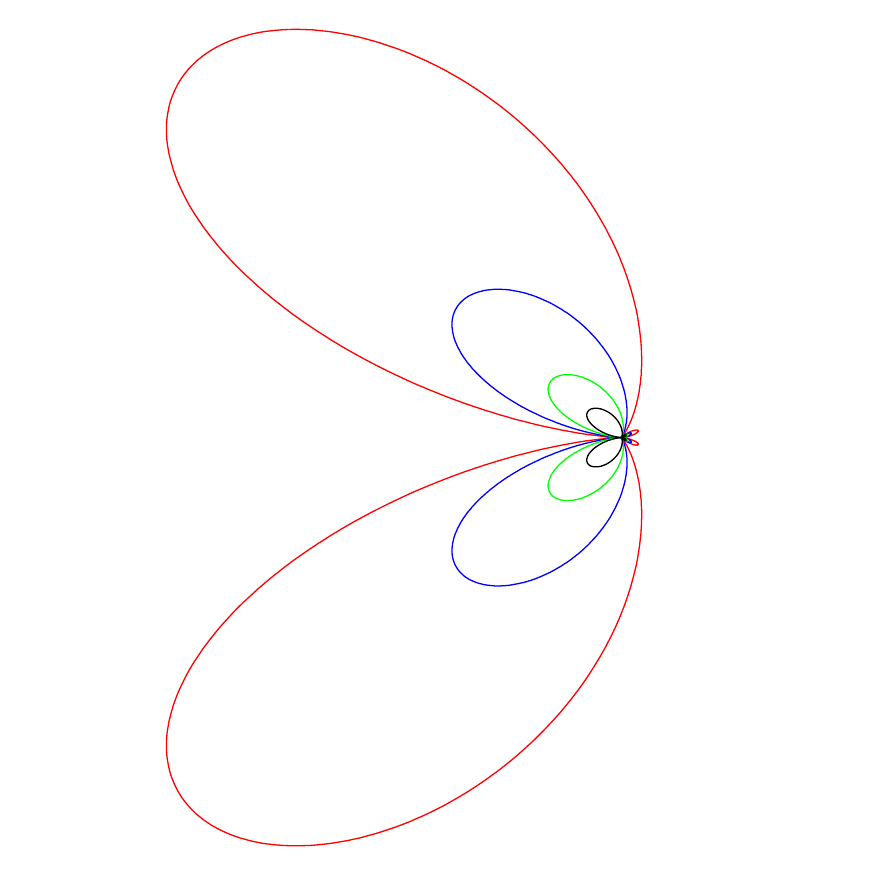} &
      \includegraphics[width=0.45\linewidth]{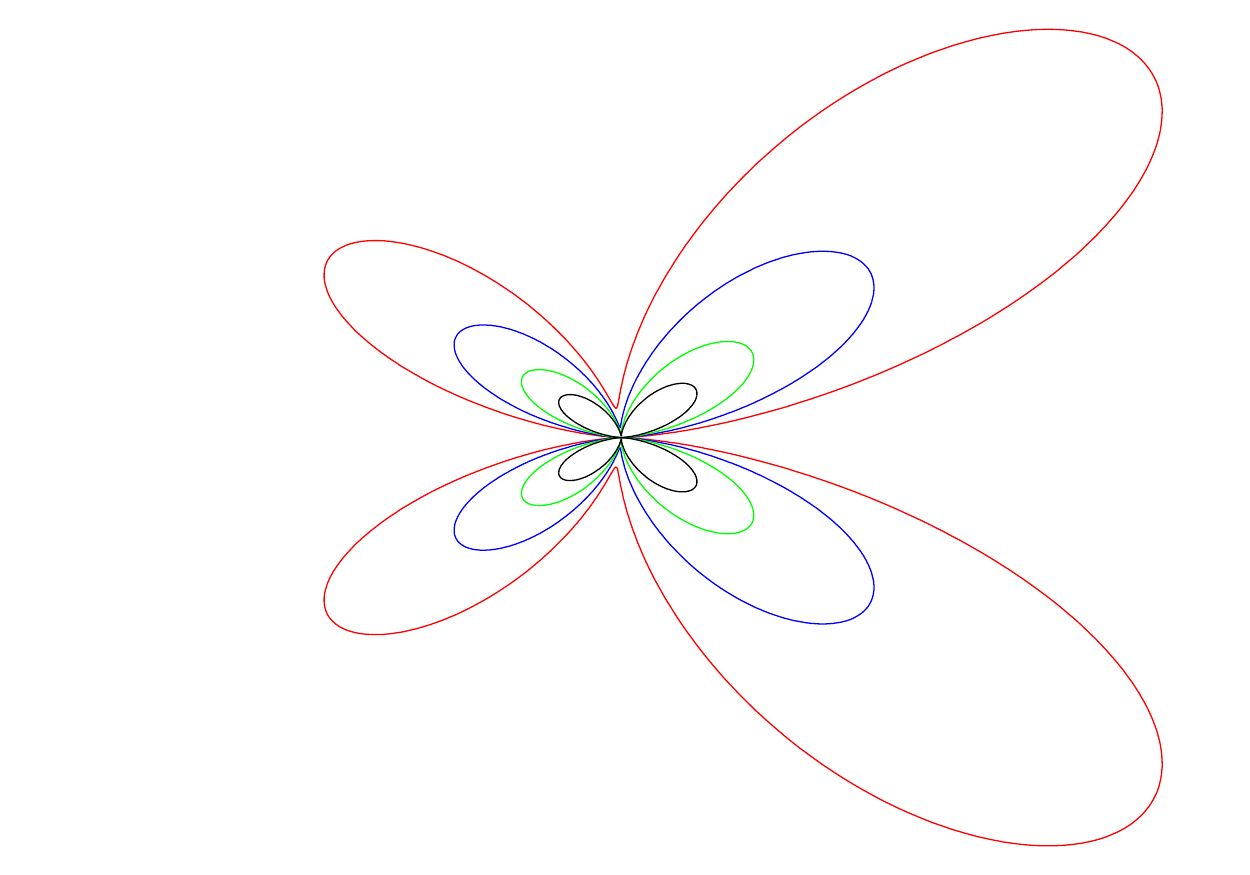}
    \end{tabular}
  \end{center}
  \caption{\label{druderad}
    Angular radiation pattern for a deformed metallic cylinder
    with deformation parameter $d=0.01$ described by a Drude response
    for frequencies $\omega$ approaching $\omega_{sp}$ or to its subharmonic:
    $\omega<\omega_{sp}/2$ (upper left),
    $\omega_{sp}/2<\omega$ (upper
    right), $\omega<\omega_{sp}$ (bottom left), and $\omega>\omega_{sp}$
    (bottom right).
    As $\omega$ approaches a resonance curves the total radiated power
    increases (color online).}
\end{figure}

In \cref{druderad} we plot the pattern $dP/d\theta$
vs. the polar angle $\theta$ corresponding to a deformed metallic
cylinder as that in \cref{drudegam}, described by the Drude response
and with a deformation parameter $d=0.01$, illuminated by a TM
electromagnetic wave propagating along the $y$ axis with an electric
field pointing along the $x$ axis, assuming that the nominal radius
$r_0$ of the cylinder is small compared to the wavelength, so we may
assume the incoming field to be constant within the particle and use
the expressions obtained in the previous section corresponding to a
homogeneous external field.
Notice there is a competition between the dipolar and quadrupolar
contributions to the radiation, and that their relative strength
varies as the frequency increases. We remark that the SH dipole would
be zero for the non-deformed cylinder, but for deformations as small
as 1\% its contribution to the radiation is comparable to the
quadrupolar contribution. \cref{druderad} shows that for low
frequencies, $\omega<\omega_{sp}/2$ displayed in the top left panel,
the radiation is completely dominated by the dipolar term and it
displays the typical pattern consisting of two symmetrical
lobes. As the frequency moves towards the resonance the total radiated
power increases hence the outermost curve in
the top left
panel corresponds to a frequency slightly below the resonance at
$\omega_{sp}/2$.  For higher frequencies the pattern
becomes largely quadrupolar. The top right
panel corresponds to $\omega>\omega_{sp}/2$ for which the quadrupolar
contribution overshadows the dipolar contribution and a
four lobed pattern emerges. It is somewhat
asymmetrical due to the intereference with the dipolar field. Also note the
shift in the size of the lobes from front to back as one moves away
from the resonance at $\omega_{sp}/2$. The bottom left panel illustrates
the radiation pattern for frequencies approaching $\omega_{sp}$ from
below and is predominantly dipolar with an influence of the
quadrupolar contribution which makes it asymmetrical. Radiation at
frequencies above that of the surface
plasmon $\omega>\omega_{sp}$ is shown in the bottom right panel. In
this region too, the pattern is mostly qudarupolar displaying four
lobes which are asymmetrical due to the interference with the dipolar
contribution to the radiation.

\begin{figure}
  \begin{center}
    \begin{tabular}{c c}
      \includegraphics[width=0.45\linewidth]{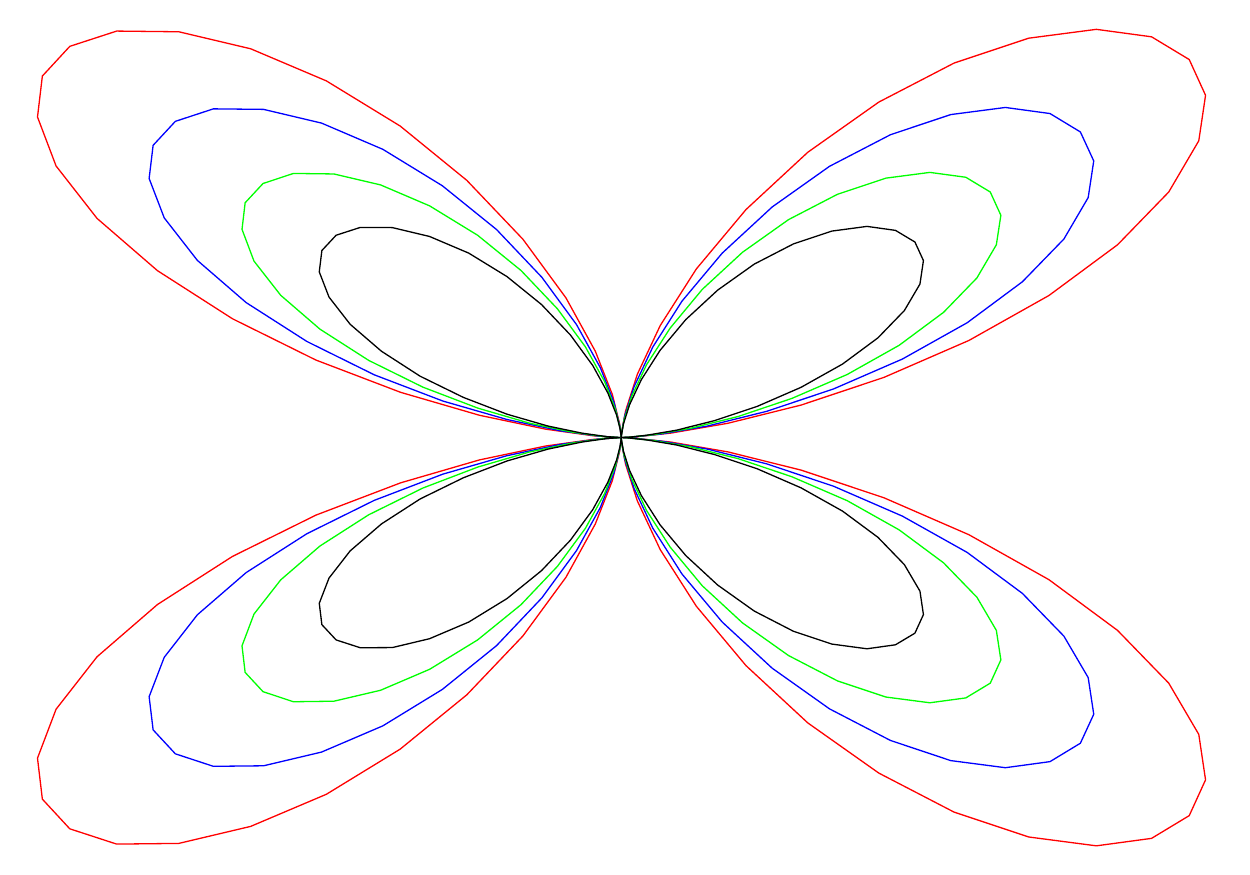} &
      \includegraphics[width=0.45\linewidth]{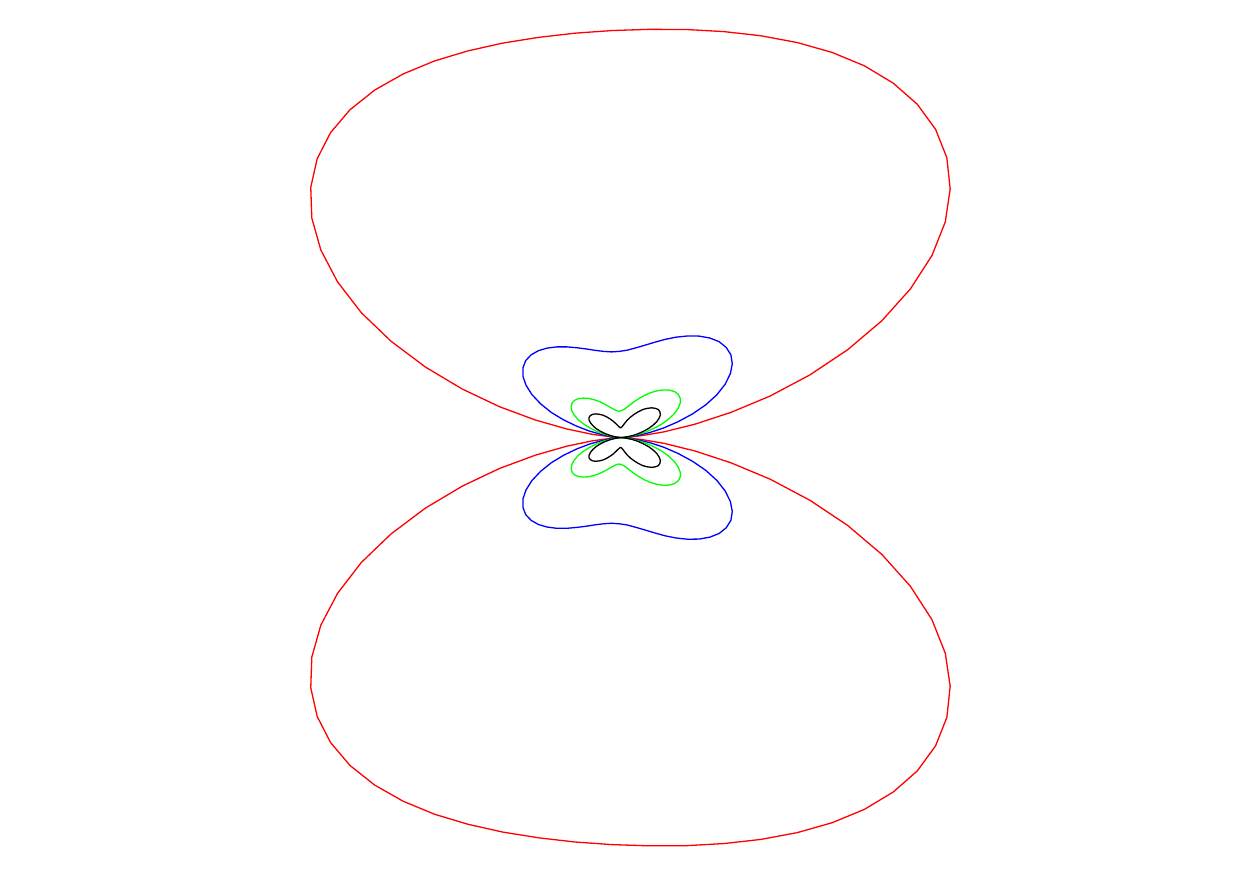} \\
      \includegraphics[width=0.45\linewidth]{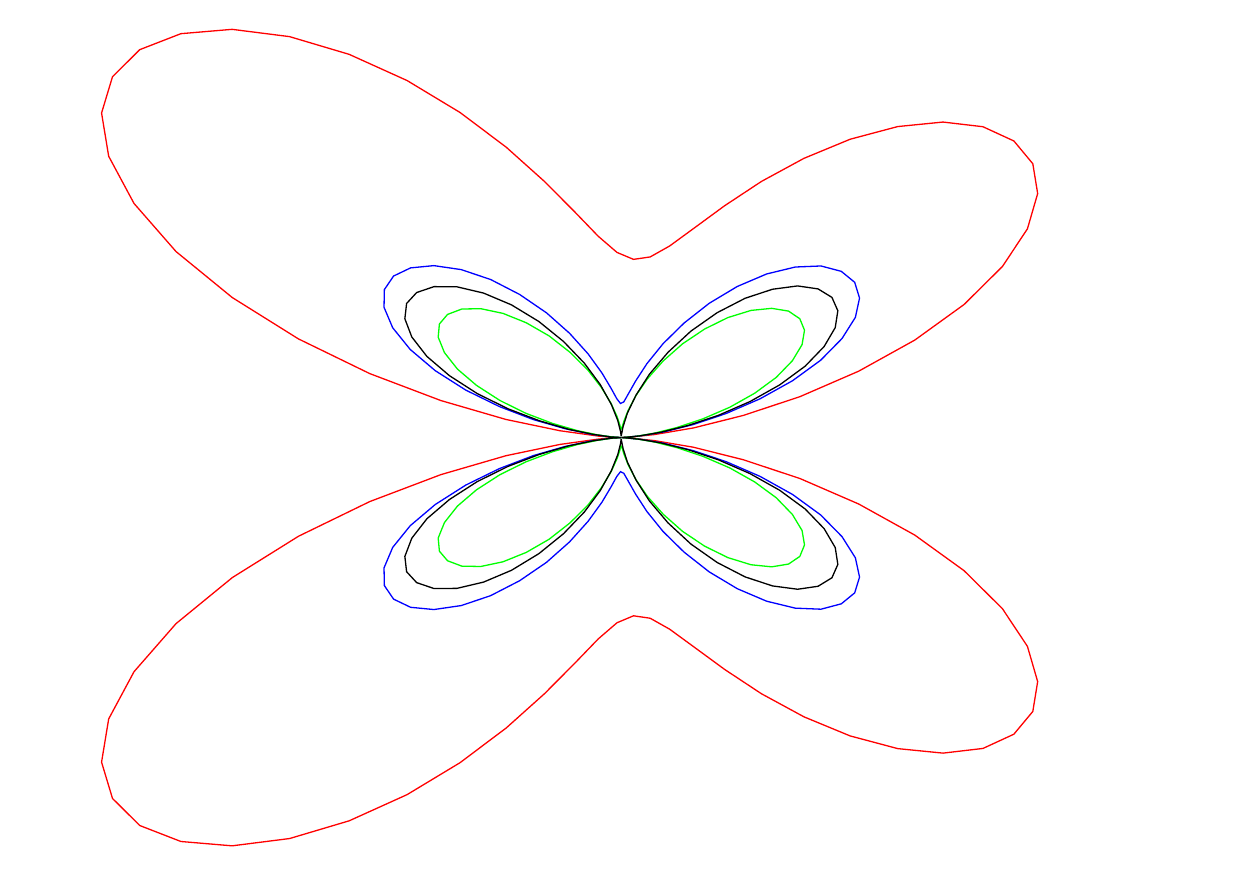} &
      \includegraphics[width=0.45\linewidth]{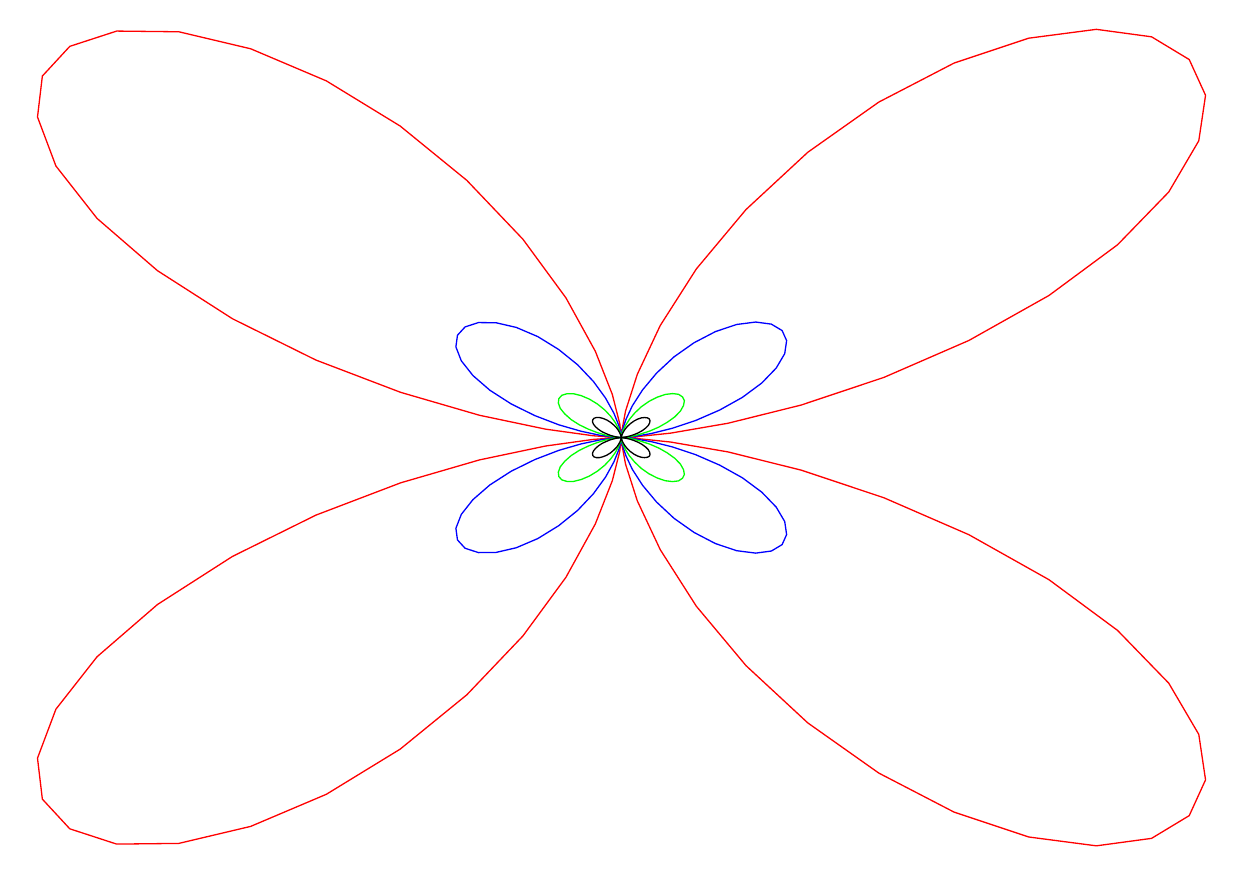}
    \end{tabular}
  \end{center}
  \caption{\label{dielrad}
    Angular radiation pattern for a deformed dielectric cylinder
    with deformation parameter $d=0.01$ described by a simple Lorentzian
    response with negligible dissipation for frequencies $\omega$
    close to $\omega_{spp}$ or its subharmonic:
    $\omega<\omega_{spp}/2$ (upper left),
    $\omega_{spp}/2<\omega$ (upper
    right), $\omega<\omega_{spp}$ (bottom left), and $\omega>\omega_{spp}$
    (bottom right).
    As $\omega$ approaches a resonance the total radiated power increases
    (color online).}
\end{figure}

In \cref{dielrad} we show the SH angular radiation pattern as in
\cref{druderad} but corresponding to a dielectric particle as in
\cref{dielgam}.
Here, we also see the competition between the dipolar and the
quadrupolar radiation with the variation in frequency and the
assymetry in the different lobes of the quadrupolar pattern arising
due to the phase difference between the two terms. Similar to
\cref{druderad}, as the frequency approaches a resonance the total
radiated power increases. However, the quadrupolar contribution
to the radiation is stronger at lower frequencies in this case unlike
the metallic case (\cref{druderad}). In the top left
panel we plot the patterns for $\omega<\omega_{spp}/2$ where the
quadrupolar contribution to the radiation overshadows the dipolar one
and is therfore symmetric. The top right panel illustrates the
radiation for frequencies $\omega>\omega_{spp}/2$. The outermost curve,
closest and slightly above the resonance at $\omega_{spp}/2$
displays a slightly distorted dipolar
pattern. Moving away from the resonance, the quadrupolar term gets
relatively stronger and the competition between the two terms gives
rise to an
asymmetry in the pattern. The bottom right panel shows the pattern for
$\omega<\omega_{spp}$ and it shows almost symmetrical quadrupolar
patterns
with the asymmetry appearing in the outermost curves, just
below the resonance frequency. The bottom right panel shows the
radiation at higher
frequencies $\omega>\omega_{spp}$ which is also symmetric and almost
pure quadrupolar like radiation.


\section{Conclusions}\label{conclusions}

We developed an analytical formalism to study the second order
nonlinear optical response of isolated particles made of
centrosymmetric materials with a cross-section slightly deformed away
from that of a centrosymmetric particle. To this end, we choose the
most simple geometry that lacks inversion symmetry, namely, a cylinder
with an almost circular cross-section with three small protuberances
separated by an angle of $2\pi/3$. We
employed a perturbative approach choosing the extent of the deformation
away from the symmetrical geometry as the smallness parameter. This
allowed us to obtain simple closed form
expressions for the electric fields within and beyond the particles
and on their surfaces at both the fundamental and second harmonic (SH)
frequencies. The
self-consistent field near the surface of the particle was used to
calculate the induced nonlinear polarization and the nonlinear
hyperpolarizabilities. The zeroeth order term in the expansion
corresponds to the case of a symmetric cylindrical particle yielding
no SH dipole but a nonzero quadrupolar response. At the first
order in the deformation the effect of a small deviation from
centrosymmetrical geometry yields a dipolar contribution proportional
to the deformation parameter which
increases with the size of deformation and competes with that of the
quadrupole already for very slightly deformed metallic and dielectric
particles. We have only considered the dominant dipolar and
quadrupolar contributions to the nonlinear hyperpolarizabilities in
this work as the higher order multipoles generate a much weaker SH
signal.

The dipolar and quadrupolar nonlinear responses were obtained
in terms of the linear dielectric response of the material at the
fundamental and SH frequencies and were found to have resonant
structures corresponding to the poles and zeroes of the dielectric
function, and to the surface plasmon or
surface plasmon-polariton frequencies of the undeformed particle, and
to their subharmonics, as
well as additional structure due to the normal nonlinear surface
parameter $a$. We showed results for particles made up of a Drude
metal and of a dielectric characterized by a simple Lorentzian
response, as they allow a simple interpretation of the
resulting spectra and radiation patterns.  Nevertheless, as the input
to our calculations are the dielectric functions of the particles,
they may be applied to particles made of arbitrary materials for which
$\epsilon$ is known. Our approach can also be generalized to other
geometries. Finally, we showed that the dipolar SH radiation
is comparable and may overshadow the quadrupolar contribution for
deformations as small
as 1\%. On the other hand, the dipolar hyperpolarizabilities (per unit
length) reached
at resonance values several orders of magnitude larger than $r_0/ne$,
with $r_0$ the nominal size of the particle, $n$ the polarizable
entity or the electronic number density and $e$ the electronic
charge (see \cref{drudegam,dielgam}). Thus, we expect that the dipolar
nonlinear susceptibility of a
metamaterial made up of these particles could be much larger than
$1/ner_0$. As the typical suceptibility of non-centrosymmetric
materials is of order $1/nea_B$ with $a_B$ the Bohr's radius, a
metamaterial made of centrosymmetric materials with a
non-centrosymmetric geometry may be a competitive source of SH provided
$a_B/r_0$ is not too small.
Finally, we remark that our calculation of the nonlinear response of
a particle showed that there are some subtleties to be accounted for:
bulk contributions, bulk
induced surface charges, surface originated
surface charges and surface dipolar layers. All of these
have to be appropriately screened to get consistent expressions for
the hyperpolarizabilities.  Analytical results for simple models that
take all of these contributions into account are important in order to
calibrate and test numerical calculations which may then, if proved to be
correct, be applied to a
larger class of systems. We
expect that the present results will be useful for this purpose.

\acknowledgments
This work was supported by DGAPA-UNAM under grants IN113016 and
IN111119 (WLM) and
by CONACyT (RS). We acknowledge useful
talks with  V. Agarwal, A. Reyes-Esqueda, L. Ju\'arez-Reyes and
B. Mendoza.

\appendix
\section*{Appendix}
\renewcommand{\theequation}{A\thesection.\arabic{equation}}

In this appendix we calculate the fields in the radiation zone due to
localized systems of oscillating charge and current densities in 2D
in order to obtain the corresponding angular radiation patterns. We
will only consider electric dipole and quadrupole radiation. The
treatment is predominantly similar to that of $3D$ \cite{jackson}, but
using the Green's function for the 2D wave equation and we follow
Ref. \onlinecite{unpublished}.

We will consider a harmonically varying monochromatic current
distribution $\bm J(\bm r,t)=\bm J(\bm r)e^{-i\omega t}$. In
the Lorentz gauge, the vector potential is also monochromatic and
obeys a wave equation which becomes a Helmholtz equation for its
amplitude $\bm A(\bm r)$ with a source $-4\pi\bm J(\bm r)/c$.
To solve it we first find the corresponding
Green's function in 2D $G(|\bm r- \bm r'|)$, which obeys
\begin{equation}
  (\nabla^{2}+k^{2})G(r)=-4\pi \delta(\bm r).
\end{equation}
Beyond the singularity, G(r)=R(kr), where
\begin{equation}
  s^2\frac{d^2}{d s^2}R(s)+s\frac{d}{d s}R(s)+
  s^2 R(s)=0.
\end{equation}
The solution is proportional to an outgoing Hankel function
$H_0^{(1)}(s)$, which in the near zone, $(s\rightarrow0)$, takes the
form
\begin{equation}\label{nearH}
  \lim_{s\rightarrow0} H_0^{(1)}(s)= 2i\log(s)/\pi.
\end{equation}
As the nonretarded Green's function in 2D is
$G=-2\ln(r)+\mathrm{constant}$, a comparison with Eq.(\ref{nearH})
yields $G(r)=i\pi H_0^{(1)}(kr)$. Thus, using the asymptotic expression
for the Hankel's function for large arguments, we obtain
in the radiation zone
$(kr\rightarrow\infty)$,
\begin{eqnarray}
  G(r)=e^{i\pi/4}\sqrt{\frac{2\pi}{kr}} e^{ikr}.
  \label{Green}
\end{eqnarray}
The retarded vector potential is then
\begin{equation}
  \bm A(\bm r)=\frac{1}{c}\int d^2r'
  e^{i\pi/4}\sqrt
  {\frac{2\pi}{k|\bm r-\bm r'|}}e^{ik|\bm r-\bm r'|} \bm J(\bm r'),
  \label{A}
\end{equation}
For a localized source in the long wavelength approximation $r'\ll
\lambda\ll r$ one can approximate Eq.(\ref{A}) as
\begin{equation}
  \bm A(\bm r)\approx\frac{1}{c}\sqrt{\frac{2\pi}{kr}}
  e^{i\pi/4}e^{ikr}\int d^2r'\, \bm J(\bm r')
  \sum_{m=0}\frac{(-ik)^m}{m!}(\hat{\bm r}\cdot\bm r')^m.
  \label{full}
\end{equation}
The first term  $(m=0)$ in the series (\ref{full}) may be integrated
to obtain the dipolar contribution to the potential,
\begin{equation}\label{dip}
  \bm A^{(0)}=(-ie^{i\pi/4})\sqrt{2\pi}\sqrt{\frac{k}{r}}e^{ikr}\bm p,
\end{equation}
where $\bm p$ is the amplitude of the oscillating dipole moment per
unit length.
The second term ($m=1$) is
\begin{equation}\label{quad1}
  \bm A^{(1)}(\bm r)=\frac{1}{c}\sqrt{\frac{2\pi}{kr}}
  e^{i\pi/4}e^{ikr}(-ik)\int d^2r'\, \bm J(\bm r')(\hat{\bm r}\cdot \bm r').
\end{equation}
Within the integral we can write
$\bm J(\bm r')(\hat{\bm r}\cdot \bm r')
=(1/2)[\bm J(\bm r')(\hat{\bm r}\cdot \bm r') +
    \bm r'(\hat{\bm r}\cdot\bm J(\bm r'))]
  +(1/2)[\bm J(\bm r')(\hat{\bm r}\cdot \bm r') -
  \bm r'(\hat{\bm r}\cdot \bm J(\bm r'))]$
as a sum of a symmetric and an antisymmetric part. The former can be
manipulated to yield
\begin{equation}\label{quad2}
  \bm A^{(1s)}(\bm r)=(\sqrt{2\pi}e^{i\pi/4})\frac{e^{ikr}}{4\sqrt{r}}k^{3/2}
  \bm Q\cdot\hat{\bm r},
\end{equation}
where $\bm Q$ is the 2D quadrupolar tensor (\cref{quad3}).
The antisymmetric part yields the magnetic dipolar radiation.

As usual, we may obtain the electromagnetic radiation field as
$\bm B=\nabla\times \bm A\approx ik\hat{\bm r}\times\bm
  A$ and $\bm E=\bm B\times \hat{\bm r}$, so that from
  \cref{dip,quad2} we obtain \cref{Brad,Erad}.

\bibliography{ref}{}
\bibliographystyle{unsrt}
\end{document}